# Neuroprotective efficacy of nimesulide against hippocampal neuronal damage following transient forebrain ischemia

Authors: Eduardo Candelario-Jalil [a, *], Dalia Álvarez [a], Armando González-Falcón [a], Michel García-Cabrera [a], Gregorio Martínez-Sánchez [a], Nelson Merino [a], Attilia Giuliani [b], Olga Sonia León [a]

Affiliation:

[a] Department of Pharmacology, University of Havana (CIEB-IFAL), Apartado Postal 6079, Havana City 10600, Cuba.

[b] Department of Chemistry and Medical Biochemistry, University of Milan, Italy.

*Author to whom all correspondence should be addressed:

**Eduardo Candelario-Jalil, M.Sc.**
**Department of Pharmacology**
**University of Havana (CIEB-IFAL)**
**Apartado Postal 6079**
**Havana City 10600**
**CUBA**
**Tel.: +53-7-271-9536**
**Fax: +53-7-336-811**
**E-mail: candelariojalil@yahoo.com**

## ABSTRACT

Cyclooxygenase-2 is involved in the inflammatory component of the ischemic cascade, playing an important role in the delayed progression of the brain damage. The present study evaluated the pharmacological effects of the selective cyclooxygenase-2 inhibitor nimesulide on delayed neuronal death of hippocampal CA1 neurons following transient global cerebral ischemia in gerbils. Administration of therapeutically relevant doses of nimesulide (3, 6 and 12 mg/kg; i.p.) 30 minutes before ischemia and at 6, 12, 24, 48 and 72 h after ischemia significantly ($P<0.01$) reduced hippocampal neuronal damage. Treatment with a single dose of nimesulide given 30 min before ischemia also resulted in a significant increase in the number of healthy neurons in the hippocampal CA1 sector 7 days after ischemia. Of interest is the finding that nimesulide rescued CA1 pyramidal neurons from ischemic death even when treatment was delayed until 24 h after ischemia ($34 \pm 9$ % protection). Neuroprotective effect of nimesulide is still evident 30 days after the ischemic episode, providing the first experimental evidence that cyclooxygenase-2 inhibitors confer a long lasting neuroprotection. Oral administration of nimesulide was also able to significantly reduce brain damage, suggesting that protective effects are independent of the route of administration. The present study confirms the ability of cyclooxygenase-2 inhibitors to reduce brain damage induced by cerebral ischemia and indicates that nimesulide can provide protection when administered for up to 24 h post-ischemia.

*Keywords:* nimesulide; cerebral ischemia; delayed neuronal death; hippocampal CA1 sector; neuroprotection; gerbils; neurodegeneration

# 1. INTRODUCTION

Transient global cerebral ischemia is produced when the brain is deprived temporarily of oxygen and glucose. In humans after cardiac arrest with resuscitation or cardiopulmonary bypass surgery, cerebral ischemia can lead to problems with cognition and memory, to serious neurological problems such as sensorimotor deficits and seizures, and to death (Levy et al., 1985; Petito et al., 1987).

In humans and in animals subjected to transient forebrain ischemia, specific neurons degenerate following the ischemic episode (Kirino, 1982, 2000; Pulsinelli et al., 1982; Petito et al., 1987). The cornu Ammonis 1 (CA1) neurons of the hippocampus are widely regarded as among the most vulnerable in the mammalian brain to ischemia (Pulsinelli et al., 1982; Kirino, 1982). Delayed hippocampal damage is observed 3 to 7 days after the insult in CA1 pyramidal neurons (Kirino, 1982), suggesting that mechanisms that develop slowly after ischemia have an important role in ischemic neuronal demise.

Several lines of recent evidences have shown that several pro-inflammatory genes or mediators, such as inducible nitric oxide synthase (iNOS), cyclooxygenase-2 and cytokines (e.g., tumor necrosis factor $\alpha$ and interleukin-1$\beta$) are strongly expressed in the ischemic brain (Ohtsuki et al., 1996; Koistinaho and Hökfelt, 1997; Barone and Feuerstein, 1999). Inflammation is now recognized as a significant contributing mechanism in cerebral ischemia because anti-inflammatory compounds or inhibitors of iNOS and cyclooxygenase-2 have been proven to reduce ischemic brain damage (Nogawa et al., 1997; Iadecola, 1997; Nakayama et al., 1998).

Nimesulide (N-(4-nitro-2-phenoxyphenyl)-methanesulfonamide) is a non-steroidal anti-inflammatory drug with potent effects, showing a high affinity and selectivity for cyclooxygenase-2 (Rabasseda, 1996; Cullen et al., 1998), but other mechanisms have been proposed to explain its mode of action: 1) inhibition of tumor necrosis factor $\alpha$ production (Azab et al., 1998), 2) antioxidant properties (Facino et al., 1993), 3) inhibition of the production of platelet activating factor (Tool and Verhoeven, 1995) and 4) reduction of the release of superoxide anions and other toxic substances from neutrophils (Bevilacqua et al., 1994). Nimesulide readily crosses the intact blood-brain barrier in both humans and rodents (Taniguchi et al., 1997; Cullen et al., 1998) .

Recently, several studies have demonstrated a marked neuroprotective effect of nimesulide on chronic cerebral hypoperfusion (Wakita et al., 1999), kainate-induced excitotoxicity (Candelario-Jalil et al., 2000) and quisqualic acid-induced neurodegeneration in rats (Scali et al., 2000).

In the light of these evidences, the present study was undertaken to investigate the effects of clinically relevant doses of nimesulide on the delayed neuronal death of CA1 pyramidal cells in the gerbil hippocampus following global ischemia. To our knowledge, there is no previous study on the effects of nimesulide against neuronal damage after focal or global cerebral ischemia.

## 2. MATERIALS AND METHODS

### 2.1. Animals and surgical procedures

Studies were performed in accordance with the Declaration of Helsinki and with the Guide for the Care and Use of Laboratory Animals as adopted and promulgated by National Institutes of Health (Bethesda, MD, USA). Our institutional animal care and use committee approved the experimental protocol (No. 01/98). A total of 111 male Mongolian gerbils (*Meriones unguiculatus*) weighing 60-75 g at the time of surgery were used in this study. These animals were housed four per cage, exposed to a 12-h light: 12-h dark cycle, and had free access to food and water throughout the study period. The gerbils were anesthetized with chloral hydrate (300 mg/kg, i.p.) and subjected to transient forebrain ischemia exactly as in our previous reports (Candelario-Jalil et al., 2001, 2002; Martínez et al., 2001). Briefly, in the supine position, a midline ventral incision was made in the neck. Both common carotid arteries were exposed, separated carefully from the vagus nerve, and occluded for 5 minutes with microaneurysmal clips (Sugita, Japan), which consistently resulted in delayed neuronal death in the CA1 region of the hippocampus (Kirino, 1982; Candelario-Jalil et al., 2002). Blood flow during the occlusion and reperfusion after removal of the clips was visually confirmed and the incision was closed with 4-0 silk sutures. In sham-operated animals, the arteries were freed from connective tissue but were not occluded. The onset of cerebral ischemia was associated with a brief period of panting breathing and body movements followed by quiescence. Successful occlusion of both common carotid arteries was evident with the rapid onset of complete bilateral ptosis and the adoption of a 'hunched' posture. Only animals showing this behavior were considered in this study (Martí et al., 2001; Martínez et al., 2001).

The rectal temperature was carefully monitored and maintained at 37 ± 0.5°C using an incandescent lamp and the animals were allowed to recover on an electrical heated blanket. In addition, rectal temperature was monitored at 6-h intervals for 3 days of reperfusion in all experimental groups.

### 2.2. Experimental protocols

#### 2.2.1. Experiment 1: Dose-response study of nimesulide treatment

This experiment was undertaken to investigate the effects of different doses of nimesulide on neuronal damage after ischemia. Since nimesulide is usually administered to patients at a dosage between 3 and 6 mg/kg for the treatment of inflammatory conditions (Rabasseda, 1996; Cullen et al., 1998), the following experimental groups were prepared: a sham-operated group (n=5), an ischemic group treated with the vehicle of nimesulide (polyvinylpyrrolidone, n=7) and three groups of ischemic gerbils treated with nimesulide at doses of 3 (n=6), 6 (n=8) and 12 mg/kg (n=7) given intraperitoneally. Nimesulide or vehicle was administered 30 minutes before ischemia and again at 6, 12, 24, 48 and 72 h of reperfusion. This treatment schedule was based on our previous experience with another cyclooxygenase-2 inhibitor (Candelario-Jalil et al., 2002). In addition, we reasoned that if cyclooxygenase-2 is one of a select few proteins that still remains upregulated in CA1 neurons even at 3 days after global ischemia (Nakayama et al., 1998; Koistinaho et al., 1999), the nimesulide treatment should be maintained for several days after the ischemic episode. We also performed an experiment in which nimesulide was given as a single dose.

### 2.2.2. Experiment 2: Effects of a single dose of nimesulide

The effect of a single dose of nimesulide (6 mg/kg; i.p.) given 30 minutes before the onset of ischemia was examined (n=7). The dose of nimesulide used in this experiment was based on results obtained from Experiment 1 (see Fig. 1). A single injection vehicle-treated group was also included (n=7).

### 2.2.3. Experiment 3: Time window for nimesulide treatment in gerbils subjected to transient forebrain ischemia

In this experiment, the effect of nimesulide was studied in a situation in which its first administration was delayed for 6 to 24 h after ischemia. Nimesulide (6 mg/kg; i.p.) was dosed 6 (n=9), 12 (n=9) and 24 h (n=7) after the induction of ischemia, followed by additional doses at 24, 48 and 72 h of reperfusion.

### 2.2.4. Experiment 4: Analysis of duration of the neuroprotective action of nimesulide

This experiment was undertaken to examine whether the protective effect of nimesulide was transitory or long lasting. Nimesulide (6 mg/kg; i.p.) was administered as in Experiment 1, but gerbils were sacrificed 30 days after the onset of blood flow (n=7). A vehicle-treated ischemic control group was also included (n=7).

*2.2.5. Experiment 5: Oral administration of nimesulide*

Since nimesulide is commonly administered to patients in oral dosage forms, in this experiment we investigated whether the protective effects of nimesulide are dependent of the route of administration. This experiment followed the exact same procedure as Experiment 1, except that nimesulide was dosed orally (6 mg/kg, n=6). In the ischemic control group (n=7) the animals were treated orally with vehicle (0.5% carboxymethylcellulose solution) for 3 days.

*2.3. Histologic analysis*

Gerbils were sacrificed 7 or 30 days (in Experiment 4) after ischemia. Animals were anesthetized with urethane (0.8 g/kg; i.p.) and perfused transcardially with cold saline followed by 4% formalin in phosphate-buffered saline (0.1 M; pH 7.4). The brains were removed from the skull and fixed in the same fixative for 24 h. Thereafter, the brains were embedded in paraffin and representative coronal sections (5-μm thick), which included the dorsal hippocampus (1.0-2.2 mm posterior to bregma), were obtained with a rotary microtome (Leica, Model RM2135, Meyer Instruments, Houston, TX, USA). Tissue sections were stained with hematoxylin and eosin. The hippocampal damage was determined by counting the number of intact neurons in the stratum pyramidale within the CA1 subfield at a magnification of 40x (Lee et al., 2000). Only neurons with normal visible nuclei were counted. The mean number of CA1 neurons per mm linear length for both hemispheres in three adjacent sections of dorsal hippocampus was calculated for each group of animals. An observer who was unaware of the drug treatment for each gerbil made all assessments of histological sections.

*2.4. Effect of nimesulide on physiologic variables*

The possible effect of nimesulide on several physiologic variables was studied in additional animals. Mean arterial blood pressure, blood glucose, hematocrit, blood pH and blood gases ($pO_2$ and $pCO_2$) were measured at baseline, during the occlusion and 30 min after each vehicle or nimesulide (12 mg/kg; i.p.) administration following the exact treatment paradigm of the Experiment 1.

*2.5. Statistical analysis*

The data of this study was expressed as a mean value ± S.D. Statistical analysis was performed using one-way analysis of variance (ANOVA) followed by Student-Newman-Keuls post-hoc test and significance refers to results where $P < 0.05$ was obtained.

## 2.6. Chemical supplies

Nimesulide was kindly provided by Gautier-Bagó Laboratories (Buenos Aires, Argentina). Polyvinylpyrrolidone and carboxymethylcellulose were obtained from BDH Chemicals (Poole, England). All other reagents were purchased from Sigma Chemical Co. (Saint Louis, MO, USA).

## 3. RESULTS

### 3.1. Effects of different doses of nimesulide on CA1 hippocampal injury (Experiment 1)

Seven days after the ischemic episode, widespread damage to the CA1 region of the hippocampus was evident in the brains of the vehicle-treated group. Pyramidal neurons either presented a densely stained shrunken appearance with minimal cytoplasm or, in many instances, had disappeared. Some of these animals displayed a nearly total necrosis or loss of the CA1 pyramidal cells.

Delayed neuronal death in CA1 hippocampal sector was significantly reduced (P<0.01) by administration of nimesulide at the three doses examined (3, 6 and 12 mg/kg; i.p.; 47 ± 3 %, 49 ± 10 % and 52 ± 9 % of protection, respectively) as shown in Fig. 1. No statistically significant difference was detected in the neuronal density between the three doses of nimesulide (Fig. 1).

Taking into account that previous studies have found marked neuroprotective effects with nimesulide at the dose of 6 mg/kg in different models of brain injury (Candelario-Jalil et al., 2000; Cernak et al., 2001), we decided to select the dose of 6 mg/kg to further evaluate the effects of nimesulide using other treatment paradigms.

### 3.2. Effects of a single dose of nimesulide

Administration of a single dose of nimesulide (6 mg/kg; i.p.) 30 minutes before ischemia resulted in a significant (P<0.05) reduction in the extent of CA1 neuronal loss as compared to the vehicle-treated group (Fig. 1), but this effect was much more evident when the treatment is continued for 3 days after the onset of reperfusion (Fig. 1), indicating that the protection depends on treatment duration.

### 3.3. Time window for nimesulide neuroprotection in gerbils subjected to 5 min transient cerebral ischemia (Experiment 3)

In this experiment, we tried to determine whether nimesulide could protect neurons when applied several hours after ischemia. To answer this question nimesulide (6 mg/kg; i.p.) was administered 6, 12 and 24 h after ischemia. All treatment schedules led to statistically significant increase (P<0.05) in the number of healthy neurons in the CA1 subfield (Fig. 2). When nimesulide administration is delayed until 6-12 h after the insult, the neuroprotection is similar (49 ± 6 and 50 ± 14 %, respectively) to that seen in the group in which the treatment started 30 minutes before ischemia (48 ± 11%), although an overall decline of efficacy with post-treatment time was observed (Fig. 2).

*3.4. Duration of the neuroprotective effect of nimesulide (Experiment 4)*

At 30 days after 5 min ischemia, brains of gerbils given vehicle or nimesulide (6 mg/kg; i.p.; for 3 days post-ischemia) were studied to determine the number of surviving neurons in the CA1 hippocampal subfield. Cell counting revealed that the neuroprotective effect of nimesulide remained virtually unchanged over the 30 days time period after the ischemic insult (Fig. 3). These data suggest that the protective effects of nimesulide are sustained over time.

*3.5. Oral administration of nimesulide*

Oral treatment with nimesulide (6 mg/kg; 30 min pretreatment) significantly attenuated (P<0.05) neuronal damage in the CA1 region of the hippocampus as evaluated by histology (vehicle 64 ± 8 neurons/mm vs nimesulide 138 ± 11 neurons/mm). The neuroprotection was similar to that obtained with intraperitoneal administration (133 ± 10 neurons/mm), indicating that the protective effect is independent of the route of administration.

*3.6. Physiologic variables in ischemic animals after nimesulide treatment*

There were no significant differences in arterial blood pressure, blood glucose, hematocrit, blood pH or blood gases between vehicle and nimesulide groups (Table 1).

## 4. DISCUSSION

The core finding of the present study is that administration of clinically relevant doses of nimesulide is remarkably neuroprotective in gerbils against transient cerebral ischemia. There is no previous report on the protective effects of this cyclooxygenase-2 inhibitor in brain ischemia. Nimesulide rescued CA1 pyramidal neurons from ischemic death even when treatment was delayed until 24 h after ischemia. Of special interest is the finding that the protection is still observed 30 days after restoration of blood flow, providing the first experimental evidence that cyclooxygenase-2 inhibitors confer a long lasting and likely permanent neuroprotection and do not merely delay cell death.

Although the neuroprotective effects of glutamate antagonists in animal models of cerebral ischemia have been remarkable (Muir and Lees, 1995), their clinical benefits are limited because the neuroprotection is not long lasting (Colbourne et al., 1999) and most of them show severe side effects.

Nimesulide has been widely used in several European countries for 20 years to treat inflammatory conditions and fever (Cullen et al., 1998). It is important to note that the neuroprotective effects of nimesulide against ischemic damage (Fig. 1) were observed at doses that are well tolerated in patients with minimal side effects.

Hypothermia during ischemia prevents ischemic injury (reviewed by Corbett and Thornhill, 2000). Unless the brain temperature is carefully controlled, neuroprotection could easily result from drug-induced hypothermia rather than a specific pharmacological effect. The protection conferred by nimesulide is not attributable to effects on body temperature because this variable was carefully monitored and did not differ between treated and non-treated groups (data not shown).

The protective time window of nimesulide treatment, until 24 h after the onset of reperfusion, is clinically exploitable in the context of cardiac arrest with resuscitation. Whereas resuscitative efforts often restore perfusion to the brain, still a considerable number of patients end up having permanent cerebral damage because only minutes of global ischemia can lead to permanent brain injury (Petito et al., 1987). Thus, if nimesulide proves as protective in humans as in rodents, resuscitation followed by nimesulide treatment would save a number of patients from permanent cerebral damage.

The inflammatory mediator pathway has been implicated as a potential contributor to the progression of ischemic injury (del Zoppo et al., 2000; Iadecola and Alexander, 2001). Cyclooxygenase-2 mRNA and protein levels have been shown to be significantly increased within neurons and vascular cells after cerebral ischemia and other insults that result in neurodegeneration (Planas et al., 1995; Nogawa et al., 1997; Sanz et al., 1997; Nakayama et al., 1998; Koistinaho et al., 1999; Iadecola et al., 1999; Dore et al., 2001). Cyclooxygenase-2 has become the focus of attention because it is the rate-limiting enzyme involved in arachidonic acid metabolism, thereby generating prostaglandins and thromboxanes, molecules that play important roles in supporting and sustaining the inflammatory response (Vane et al., 1998).

Our present result that nimesulide protects CA1 neurons when administered several hours after ischemia (Fig. 2) is consistent with previous studies which have found that cyclooxygenase-2 selective inhibitors have a wide therapeutic window for protection in both focal (Nogawa et al., 1997) and global cerebral ischemia (Candelario-Jalil et al., 2002). A similar long therapeutic time window in brain ischemia has been reported for other pharmacological strategies aimed at reducing post-ischemic inflammation (Iadecola et al., 1995; Nagayama et al., 1998; Cash et al., 2001). In addition, several studies have found neuroprotective effects of different compounds when administered after ischemia (Jiang et al., 1998; Yrjanheikki et al., 1999; Haag et al., 2000; Phillips et al., 2000; Zhang et al., 2001; Snider et al., 2001; Lavie et al., 2001; Mary et al., 2001), suggesting that brain injury produced by cerebral ischemia develops over a period of hours to days after the primary event.

Repeated treatments with nimesulide for the first three days after ischemia afforded a more remarkable neuroprotection than the administration of a single dose given before the insult (Fig. 1). These data show the importance of continuous long-term administration after ischemic damage in clinical trials to achieve the maximal beneficial effects of neuroprotection by nimesulide to target the delayed progression of damage.

Increased cyclooxygenase-2 activity and enhanced release of prostanoids have been shown to be associated with the generation of highly reactive oxygen species, which have potent deleterious effects on cells. Recently, we have found that administration of the cyclooxygenase-2 inhibitor nimesulide at a dose proven to confer protection in the present study (6 mg/kg), is able to reduce hippocampal oxidative damage following excitotoxic brain injury (Candelario-Jalil et al., 2000), which is a key pathological event of cerebral ischemia. This latter observation is particularly relevant in view of evidence showing that oxidative stress is implicated in the development and progression of apoptotic cell death in the central nervous system (Sastry and Rao, 2000) and that activation of cyclooxygenase-2 is required for execution of oxidative neuronal death (Lee et al., 2001). Further experiments are needed to clarify the effects of selective inhibition of cyclooxygenase-2 on markers of oxidative stress in the ischemic brain and its relation to neuronal death after ischemia.

Several additional mechanisms could account for the neuroprotection conferred by nimesulide in global cerebral ischemia. The possibility that nimesulide diminished neuronal loss through reduction of oxidative damage, inhibition of pro-inflammatory cytokines production and blockade of apoptotic pathways cannot be excluded.

In summary, the present study has demonstrated a marked neuroprotective effect of nimesulide on the gerbil CA1 hippocampal neurons after transient forebrain ischemia at therapeutically relevant doses when administered even 24 h after the initial ischemic episode. Of great importance is the finding that neuroprotection conferred by nimesulide is long lasting and also observed with oral treatment. Our results suggest that cyclooxygenase-2 selective inhibition by nimesulide may be of therapeutic benefit in the treatment of global cerebral ischemia.

**Table 1.** Physiologic variables in additional gerbils (n=6 per group) given nimesulide (12 mg/kg; i.p.)

30 min before carotid occlusion and at 6, 12, 24, 48 and 72 h of reperfusion.

| Time (h) | Treatment | Blood pressure (mmHg) | Blood glucose (mmol/L) | Hematocrit (%) | pH | pO$_2$ (mmHg) | pCO$_2$ (mmHg) |
|---|---|---|---|---|---|---|---|
| Baseline | Vehicle | 69 ± 5 | 5.0 ± 0.6 | 47.9 ± 0.9 | 7.33 ± 0.03 | 128 ± 24 | 34.8 ± 6.5 |
| | Nimesulide | 67 ± 6 | 5.3 ± 0.8 | 49.3 ± 0.5 | 7.35 ± 0.04 | 132 ± 19 | 32.5 ± 5.8 |
| During Occlusion | Vehicle | 86 ± 8 | 4.9 ± 1.4 | 48.0 ± 0.7 | 7.31 ± 0.05 | 107 ± 41 | 26.8 ± 7.8 |
| | Nimesulide | 82 ± 11 | 5.2 ± 0.7 | 49.1 ± 0.9 | 7.24 ± 0.02 | 114 ± 15 | 28.4 ± 6.9 |
| 30 min after reperfusion | Vehicle | 57 ± 15 | 4.8 ± 0.6 | 47.2 ± 0.6 | 7.29 ± 0.06 | 156 ± 22 | 23.4 ± 8.9 |
| | Nimesulide | 59 ± 19 | 5.1 ± 1.1 | 49.2 ± 0.4 | 7.26 ± 0.07 | 153 ± 16 | 25.2 ± 3.7 |
| 6 h 30 min | Vehicle | 62 ± 12 | 5.6 ± 0.7 | 48.3 ± 0.5 | | | |
| | Nimesulide | 69 ± 9 | 5.0 ± 0.8 | 47.5 ± 0.7 | | | |
| 12 h 30 min | Vehicle | 70 ± 14 | 4.1 ± 0.9 | 48.7 ± 0.4 | | | |
| | Nimesulide | 62 ± 9 | 5.2 ± 1.0 | 49.1 ± 0.8 | | | |
| 24 h 30 min | Vehicle | 58 ± 13 | 4.0 ± 1.5 | 48.2 ± 0.5 | | | |
| | Nimesulide | 64 ± 9 | 4.7 ± 0.8 | 48.5 ± 0.6 | | | |
| 48 h 30 min | Vehicle | 69 ± 8 | 4.8 ± 0.4 | 48.6 ± 0.5 | | | |
| | Nimesulide | 70 ± 13 | 5.3 ± 0.7 | 47.3 ± 0.4 | | | |
| 72 h 30 min | Vehicle | 66 ± 12 | 4.8 ± 1.2 | 48.8 ± 0.2 | | | |
| | Nimesulide | 67 ± 5 | 5.2 ± 1.4 | 49.2 ± 0.5 | | | |

Values are mean ± S.D. Monitoring of physiological variables was performed at 30 min after nimesulide or vehicle injection. No statistically significant differences were found between gerbils treated with vehicle or nimesulide (n=6 per group) (P>0.05, Student's *t* test).

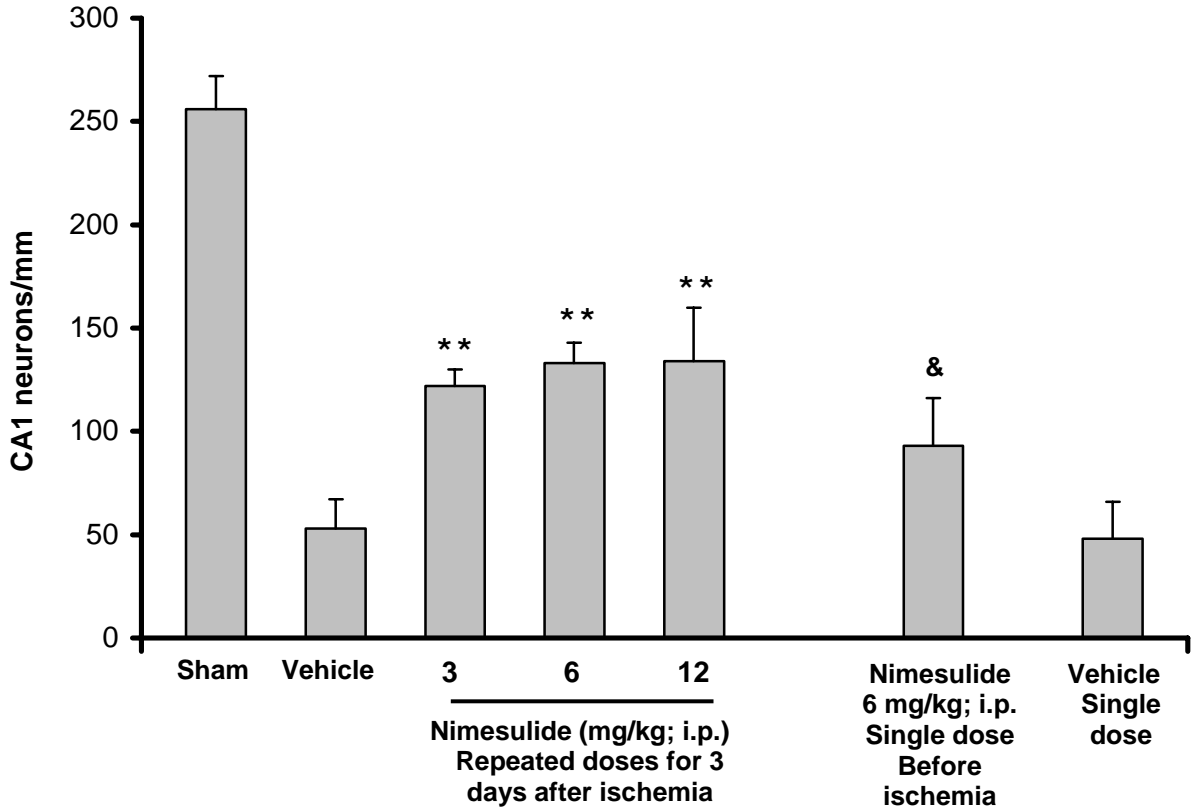

**Fig. 1.** Effect of clinically relevant doses (3, 6 and 12 mg/kg; i.p.) of the selective cyclooxygenase-2 inhibitor nimesulide on the number of surviving neurons in the CA1 hippocampal subfield 7 days after 5 min transient global cerebral ischemia in Mongolian gerbils. Repeated treatment with nimesulide for 3 days after ischemia (30 min before ischemia and after 6, 12, 24, 48 and 72 h of reperfusion) afforded a more remarkable neuroprotective effect than the administration of a single dose (6 mg/kg; i.p.) given 30 min before ischemia. Values are mean counts of normal-appearing CA1 neurons per mm linear length ± S.D. *P<0.05 and **P<0.01 for the comparison between nimesulide-treated and vehicle-treated groups. [&]P<0.05 with respect to single injection vehicle-treated group. (One-way ANOVA followed by Student-Newman-Keuls post-hoc test).

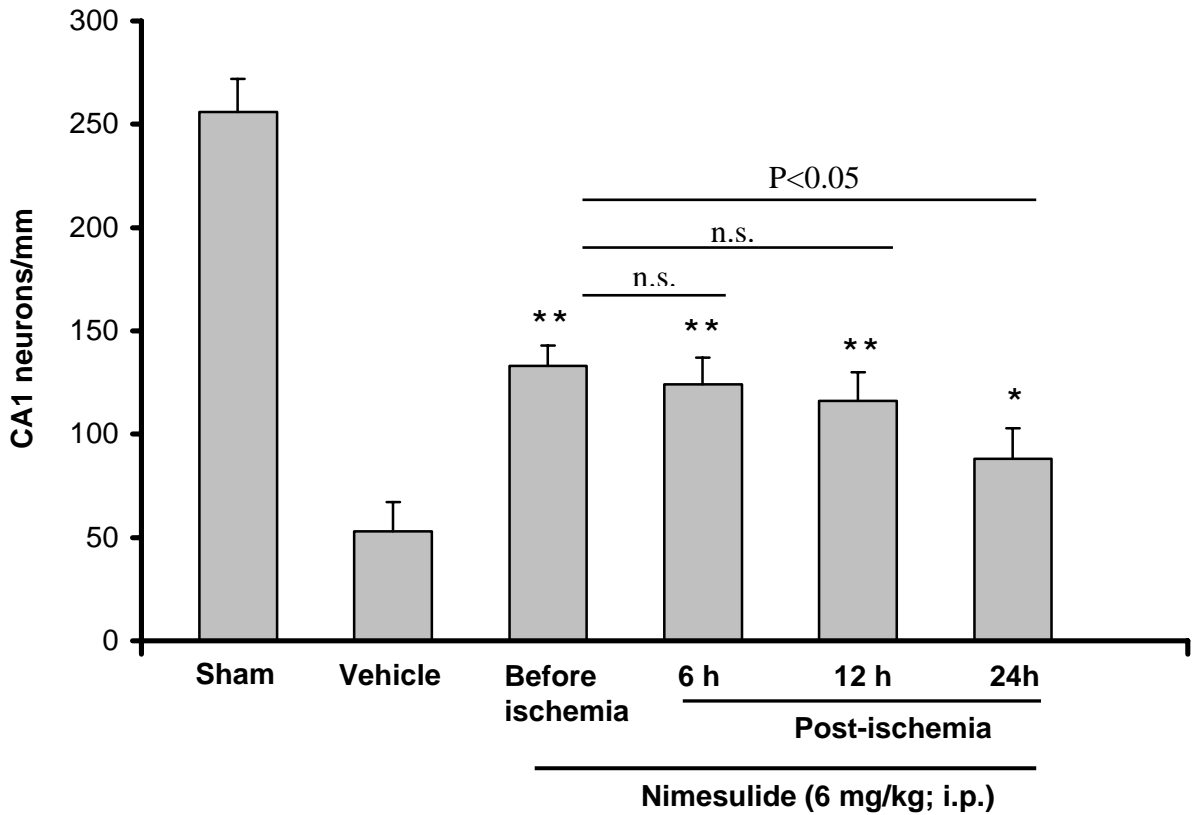

**Fig. 2.** Time-dependency of the neuroprotective effect of nimesulide on 5 min of transient global ischemia. The drug (6 mg/kg; i.p.) was administered 6 (n=9), 12 (n=9) and 24 hours (n=7) after restoration of blood flow. When nimesulide treatment is delayed until 12 h after ischemia, neuroprotection is similar to gerbils pretreated with the drug. Data are always expressed as healthy neurons per mm linear length of the hippocampal CA1 region ± S.D. *P<0.05 and **P<0.01 with respect to vehicle-treated group (ANOVA followed by Student-Newman-Keuls multiple comparison test).

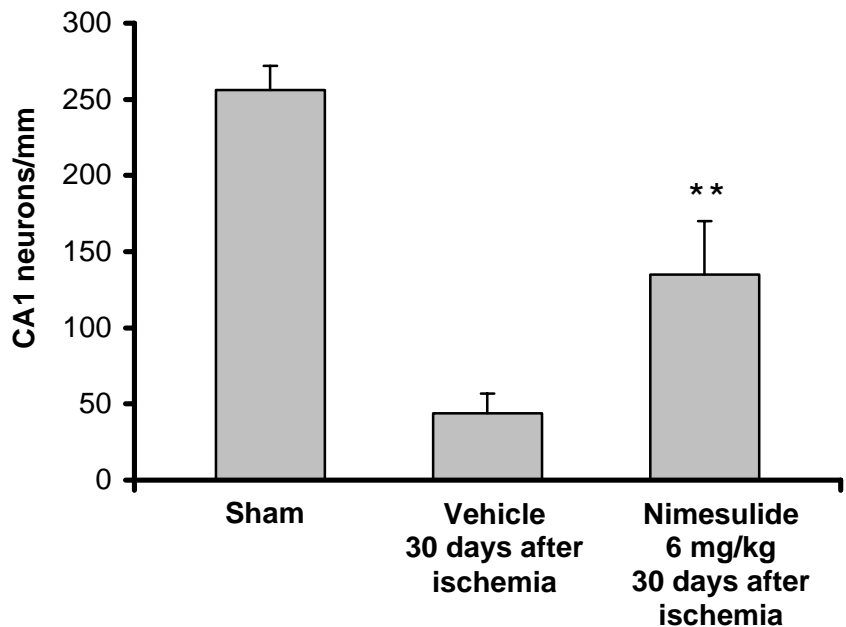

**Fig. 3.** Effect of nimesulide (6 mg/kg; i.p., for 3 days post-ischemia) on the number of surviving neurons in the CA1 hippocampal region 30 days after the ischemic episode. Cell counting revealed that the neuroprotective effect of nimesulide remained virtually unchanged over the 30 days time period after ischemia. Values are mean counts of healthy CA1 neurons per mm linear length ± S.D. **$P<0.01$ with respect to vehicle-treated animals. One-way ANOVA followed by Student-Newman-Keuls multiple comparison test.